\documentclass[aip,apl,reprint]{revtex4-1}
 
\usepackage[]{graphicx}
\usepackage{graphics}
\usepackage{amsmath}
\usepackage{amsfonts}
\usepackage{amssymb}
\usepackage{wasysym} 
\usepackage{nicefrac} 
\usepackage{setspace}
\usepackage{color} 
\usepackage{nicefrac} 
\usepackage{dcolumn}%
\usepackage{bm}%

\renewcommand{\Im}{\operatorname{Im}}

\newcommand{\figref}[1]{Fig.~\ref{fig:#1}}
\newcommand{\Figref}[1]{Figure~\ref{fig:#1}}

\renewcommand{\eqref}[1]{Eq.~(\ref{eq:#1})}

\newcommand{\citeasnoun}[1]{Ref.~\onlinecite{#1}}

\begin{document}

\preprint{AIP/123-QED}

\title{Thermal bistability through coupled photonic resonances}

\author{Chinmay Khandekar}
\author{Alejandro W. Rodriguez} 
\address{Princeton University, Princeton, NJ 08544, USA}


\begin{abstract}
  We present a scheme for achieving thermal bistability based on the
  selective coupling of three optical resonances. This approach
  requires one of the resonant frequencies to be temperature
  dependent, which can occur in materials exhibiting strong
  thermo-optic effects. For illustration, we explore thermal
  bistability in two different passive systems, involving either a
  periodic array of Si ring resonators or parallel GaAs thin films
  separated by vacuum and exchanging heat in the near field. Such a
  scheme could prove useful for thermal memory devices operating with
  transition times $\lesssim$ hundreds of milliseconds.
\end{abstract}



\maketitle

Rapid progress in the synthesis and processing of materials at small
lengthscales has created demand for understanding thermal phenomena in
nanoscale systems.~\cite{cahill2003nanoscale,li2012colloquium} Recent
interest in harnessing excess heat that is readily available at the
nanoscale has culminated in several proposed thermal
devices~\cite{song2015near} with various functionalities, including
thermal rectifiers,~\cite{otey2010thermal} thermal
memory,~\cite{wang2008thermal} thermal transistors,~\cite{ben2014near}
phononic logic gates,~\cite{sklan2015splash} and phonon
waveguides.~\cite{chang2007nanotube}. In this paper, we propose a
scheme to achieve thermal bistability based on the coupling between
three or more optical resonances. Our approach complements and builds
on recently proposed ideas~\cite{kubytskyi2014radiative,
  dyakov2015near,wang2008thermal,zhu2012negative} in several ways,
described further below.

A thermal, bistable system can be used as a memory device that stores
thermal information by maintaining the temperature of the system in
one of two or more possible states. Realizing such temperature
bistability requires a nonequilibrium thermal circuit supporting
multiple steady states. Such a circuit was first proposed several
years ago based on the concept of negative differential thermal
resistance (NDTR), which relies on the ability to achieve heat flux
rates between objects that decrease with increasing temperature
differences. While first proposed in a model system consisting of a
lattice of one-dimensional nonlinear mechanical
oscillators,~\cite{wang2008thermal} recent implementations of NDTR
have instead sought to exploit radiative energy transfer between slabs
separated by nanometer gaps and heated to very high $\approx 1500$K
temperatures.~\cite{zhu2012negative,elzouka2014near}. Here, we propose
a simple and experimentally feasible, all-optical scheme based on a
system of three optical resonances that builds and expands on a
recently proposed and related scheme which requires materials
supporting metal-insulator
phase-transitions.~\cite{kubytskyi2014radiative,dyakov2015near}
Instead, our approach exploits common materials exhibiting strong
thermo-optic effects and relies instead on thermal bistability induced
by a resonant mechanism involving three optical resonances---microring
cavities supporting travelling-wave resonances or polar--dielectric
slabs supporting surface--propagating polaritonic resonances. This
work extends previous studies of thermal
rectification~\cite{chen2014photon,otey2010thermal} and NDTR through
vacuum~\cite{zhu2012negative} and also parallels recent ideas based on
exotic non-volatile memory
systems,~\cite{elzouka2014near,maghsoudi2016thermally,rueckes2000carbon}
which have recently been proposed as viable alternatives to
traditional electrostatic
memory.~\cite{nguyen1999radiation,bagatin2017space}

\begin{figure}[t!]
\centering
\includegraphics[width=1\linewidth]{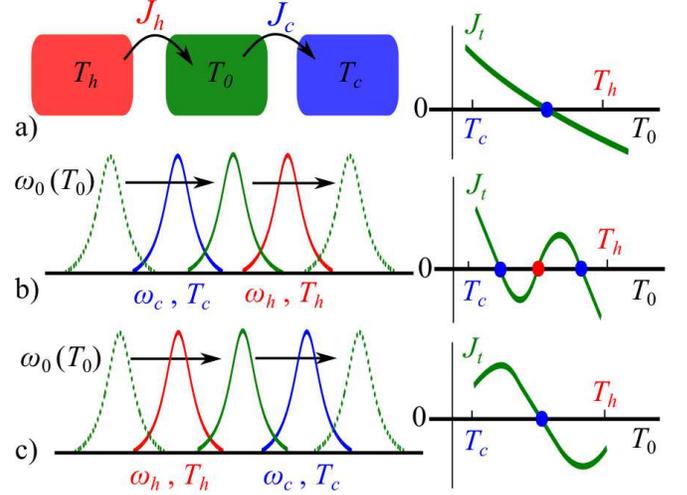}
\caption{(a) Schematic of three bodies in a nonequilibrium
  configuration of temperatures $T_h>T_0>T_c$, where the hot and cold
  bodies are maintained at constant temperatures $T_h$ and $T_c$,
  respectively, while the middle body has variable temperature
  $T_0$. The system reaches a steady state when the net heat exchanged
  $J_t=J_h-J_c=0$ (blue dot), where $J_h$ and $J_c$ denote the heat
  flux rates between the hot and cold bodies and the middle body. When
  the main mechanism behind heat exchange is a resonant process
  mediated by photonic resonances in the bodies, the variation of
  $J_t(T_0)$ can become nonmonotonic. The schematics in (b,c)
  illustrate the expected dependence of $J_t$ with $T_0$ when the
  relative frequencies of the resonators are either (b)
  $\omega_0(T_c)<\omega_c<\omega_h$ or (c)
  $\omega_0(T_c)<\omega_h<\omega_c$, where $\omega_0(T_0)$ depends
  linearly with $T_0$ due to thermoptic effects, illustrating the
  existence of multiple (stable or unstable) steady states.}
\label{fig:mech}
\end{figure}

\emph{Thermal bistability in triply resonant structures.---} We begin
by briefly describing the main mechanism behind the proposed thermal
bistability scheme, leaving quantitative predictions for later.
Consider a system of three thermal bodies, shown schematically in
\figref{mech}(a), two of which are maintained at constant temperatures
$T_h$ and $T_c$, with $T_h > T_c$, while the remaining body is
thermally isolated from its surroundings and has variable temperature
$T_0$. The hot and cold bodies exchange heat with the isolated body
through flux rates $J_h$ and $J_c$, respectively, leading to a net
heat influx $J_t=J_h-J_c$ and a steady-state temperature satisfying
(neglecting losses due to thermal conduction) $\rho c_p V
\frac{\partial T_0}{\partial t}=J_t=0$, where $\rho, c_p, V$ are the
density, specific heat capacity, and volume of the body,
respectively. For typical heat transfer mechanisms such as
conduction~\cite{holman2001heat} or
radiation,~\cite{rytov1988principles} the heat flux between any two
bodies increases with increasing temperature difference, leading to
monotonic $J_t(T_0)$ and thereby giving rise to a single steady state,
i.e. $J_t(T_0)=0$, as illustrated in \figref{mech}(a). As recently
illustrated in~\citeasnoun{zhu2012negative}, NDTR can be realized in
the context of radiative heat transfer between bodies exhibiting
significant thermo-optic effects: namely, by exploiting the monotonic
increase in the frequency of planar resonances with increasing
temperature. Here we extend this idea by considering a system of three
bodies that support narrow and slightly detuned resonances of
frequencies $\omega_j$, with $j\in\{h,0,c\}$. Consider a situation
under which $T_0=T_c$ and $\omega_0<\omega_c<\omega_h$. As the
temperature $T_0$ is increased from $T_c \to T_h$, $\omega_0$ sweeps
over the frequencies of both the hot and cold resonators, whose
temperatures and frequencies are held fixed. As $\omega_0\to
\omega_c$, the two resonators exchange heat more effectively and hence
experience larger overall heat loss, causing $J_t$ to decrease
considerably. As $\omega_0$ moves past $\omega_c$ and approaches
$\omega_h$, $J_t$ increases again due to increased coupling with the
hot resonance, decreasing with increasing $\omega_0$ as it moves past
$\omega_h$. Thus, if properly engineered, such a system can lead to
three steady states, consistent with zero net heat exchange
($J_t=0$). Such a situation is illustrated on the right half of
\figref{mech}(b), wherein the intermediate state (red dot) is unstable
while the remaining two (blue dots) are stable, i.e. $\frac{\partial
  J_t}{\partial T_0} < 0$. If, on the other hand, the initial
configuration is such that $\omega_0<\omega_h<\omega_c$ when
$T_0=T_c$, similar arguments imply the existence of a single steady
state, as illustrated in \figref{mech}(c). While this NDTR scheme can
be generalized to any system of resonances, below we consider and
quantify the feasibility of observing thermal bistability using this
scheme in realizations based on Si photonic resonators and GaAs thin
films exchanging heat in the near field. Note that nonlinear
thermo-optic effects in driven photonic resonators have been shown to
lead to optical bistability~\cite{zhang2013multibistability}, but
their use as ultrafast optical memory devalues their potential as a
slow thermal memory. In this work, we focus on passive systems in line
with previous implementations of thermal memory, in which case no
optical driving mechanisms are employed.

\begin{figure}[t!]
\centering
\includegraphics[width=0.9\linewidth]{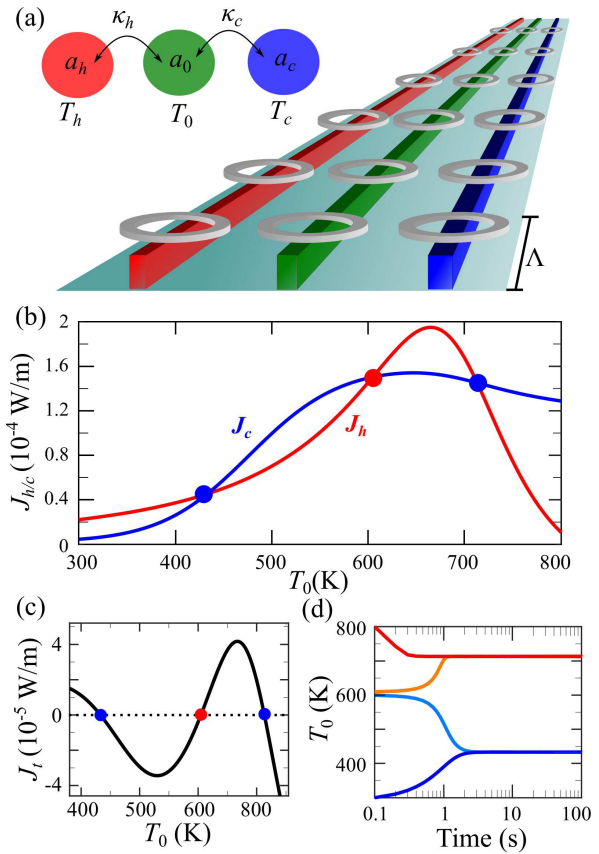}
\caption{(a) Schematic showing three arrays of Si ring resonators with
  periodicity $\Lambda=1\mu\mathrm{m}$, two of which are maintained at
  temperatures $T_h=800K$ (red) and $T_c=300K$ (blue), while the
  middle one (green) is suspended on insulating posts and has variable
  $T_0$. The heat-exchange rates per unit length between the hot/cold
  and intermediate arrays (unit cell shown schematically in the top
  inset) are $J_h$ and $J_c$, respectively, plotted in (b) as a
  function of $T_0$ under suitable operating parameters (see
  text). (c) The nonmonotonicity in the net flux rate $J_t(T_0) =
  J_h-J_c$ indicates the existence of three steady states, two of
  which are stable (blue dots) and one unstable (red dot). (d)
  Temporal relaxation of the intermediate resonators starting from
  either $T_h$ (red) or $T_c$ (blue), along with the value of the
  unstable temperature point $T_u$, as they reach the steady-state
  temperatures, $T_{s1}$ and $T_{s2}$.}
\label{fig:figSi}
\end{figure}

\emph{Ring resonators:} We first focus on a system of Si ring
resonators,~\cite{bogaerts2012silicon,xia2013suspended} which exhibit
both large thermo-optic coefficients and long-lived resonances at mid
infrared wavelengths $\sim$ peak thermal wavelength $\lambda_T \sim
10\mu m$. In particular, we consider three one dimensional arrays of
ring resonators shown in \figref{figSi}(a) with period $\Lambda$, two
of which are maintained at fixed $T_h=800$K (left) and $T_c=300$K
(right) while the middle one is suspended on insulating posts and has
variable temperature $T_0$. We ignore the negligible interactions
between adjacent rings along the array ensured by a sufficiently large
$\Lambda$ and obtain the flux rates by considering heat exchange for
three coupled resonators shown schematically in the top inset of
\figref{figSi}(a). Such a simplified system can be described via the
temporal coupled-mode theory
framework,~\cite{suh2004temporal,khandekar2015thermal,zhu2013temporal}
which provides accurate predictions while circumventing the need for
numerically intensive calculations~\cite{Hausbook}. In this framework,
the resonances are described by mode amplitudes $a_j$, normalized such
that $|a_j|^2$ are mode energies,~\cite{Hausbook} and have frequencies
$\omega_j$ and decay/loss rates $\gamma_j$, where $j=\{h,c,0\}$. They
are subject to thermal noise sources $\xi_j$ described by
delta-correlated, Gaussian noise terms satisfying $\langle
\xi_j^*(\omega)\xi_j(\omega')\rangle=
\delta(\omega-\omega')\Theta(\omega_j,T_j)$ where $\langle \cdots
\rangle$ denotes the statistical ensemble average and
$\Theta(\omega,T)=\hbar\omega/(e^{\frac{\hbar\omega}{k_B T}}-1)$ is
the Planck function. The three resonators are coupled to one another
via coupling coefficients $\kappa_h$ and $\kappa_c$, allowing heat to
flow from the hot to the cold resonator as described by the following
coupled-mode equations:
\begin{align}
\frac{da_h}{dt} &= i\omega_h a_h -\gamma_h a_h + i\kappa_h a_0 +
\sqrt{2\gamma_h}\xi_h \label{eq:eqah}\\ \frac{da_0}{dt} &= i\omega_0(T_0) a_0
-\gamma_0 a_0 + i\kappa_h a_h + i\kappa_c a_c + \sqrt{2\gamma_0}\xi_0
\label{eq:eqa0}
\\ \frac{da_c}{dt} &= i\omega_c a_c -\gamma_c a_c + i\kappa_c a_0 +
\sqrt{2\gamma_c}\xi_c \label{eq:eqac} 
\end{align}
Here, $\omega_0$ depends on the local resonator temperature through
the thermo-optic effect,~\cite{bogaerts2012silicon} with
$\omega_0(T_0) \approx \omega_0(T_c)+\frac{\omega_0}{n}\frac{\partial
  n}{\partial T_0}(T_0-T_c)$, where $n$ is the effective refractive
index and $\frac{\partial n}{\partial T_0}$ the themo-optic
coefficient of the resonator. It follows that the spectral flux
densities associated with the coupled modes, $\Phi_{h/c} =
2\Im\{\kappa_{h/c}\} \langle a^*_{h/c} a_0\rangle$, are given by:
\begin{align}
\Phi_h(\omega)&=\frac{4\kappa_h^2\gamma_h(\kappa_c^2\gamma_c
  \Theta_{hc}(\omega)+
  \gamma_0|D_c(\omega)|^2\Theta_{h0}(\omega))}{|D_h(\omega)
  D_c(\omega) D_t(\omega)|^2}
\label{eq:Phih}
\\ \Phi_c(\omega)&=\frac{4\kappa_c^2\gamma_c(\kappa_h^2\gamma_h\Theta_{hc}(\omega)
  +\gamma_0|D_h(\omega)|^2\Theta_{0c}(\omega))}{|D_h(\omega)
  D_c(\omega) D_t(\omega)|^2}
\label{eq:Phic}
\end{align}
where $\Theta_{jk}(\omega)=\Theta(\omega,T_j)-\Theta(\omega,T_k)$ and
$D_j(\omega) = i(\omega-\omega_j)+\gamma_j$, for $j,k \in \{h,0,c\}$,
and
\begin{align*}
D_t(\omega) =
 D_0(\omega)+\frac{\kappa_h^2}{D_h(\omega)}+\frac{\kappa_c^2}{D_c(\omega)}.
\end{align*}
The net flux rates per unit length, $J_{h/c}=\frac{1}{\Lambda}\int
\Phi_{h/c}(\omega) \frac{d\omega}{2\pi}$, are obtained by integrating
over all frequencies.

As an illustration, we consider rings of radii
$R_j=\frac{c}{n\omega_j}$ for $j\in \{h,0,c\}$, designed to support
resonances at $\omega_c=3.62\times 10^{14}$rad/s ($5.2\mu$m),
$\omega_h=\omega_c+7\gamma_c$, and $\omega_0(T_c)=\omega_c-2\gamma_c$,
with equal decay rates $\gamma_j=\omega_c/500$. Material properties
are thermo-optic coefficient $\frac{\partial n}{\partial T}=2\times
10^{-4}K^{-1}$ and effective refractive index $n=3.42$. With ring
radii $R_j\sim 0.25\mu\mathrm{m}$, lattice period
$\Lambda=1\mu\mathrm{m}$ is chosen to ignore the interactions between
neighboring rings along the array and the arrays are placed such that
the coupling rates $\kappa_h=0.9\gamma_c$ and
$\kappa_c=2\gamma_c$. \Figref{figSi}(b) shows the flux rates $J_h$ and
$J_c$ per unit length as a function of the temperature $T_0$ of the
middle resonator. The net flux entering/leaving the ring $J_t$ shown
in \figref{figSi}(c) leads to two stable steady states at
$T_{s1}=413K$ and $T_{s2}=700K$ (blue dots), along with an unstable
state at $T_{u}=600K$ (red dot). Here, we ignore radiative decay into
the surroundings as well as conductive losses into the mechanically
supporting structures. These extraneous channels of heat transfer can
be suppressed by suspending the middle rings on thermally insulating
posts to reduce conductive losses (see bottom schematic) while also
operating under vacuum to eliminate conductive/convective heat
transfer through air, as discussed in
\citeasnoun{shaw2008fabrication}. Apart from stability against
temperature perturbations, guaranteed here by large temperature gaps
between steady states, robustness against flux perturbations will
generally depend on the flux barrier and hence net magnitude of the
flux rates $\sim \Theta(\omega,T)/Q$, guaranteed here by operating
with large wavelengths and relatively small $Q$. \Figref{figSi}(d)
illustrates the relaxation of $T_0$ from $T_h,T_c,T_u^+,T_u^-$ to the
nearest stable steady states $T_{s1},T_{s2}$, assuming $V\sim 0.1
\mu$m$^3$ where $V$ is the volume of the middle ring and the
temperature-dependent values of $c_p$ and $\rho$ given
in~\citeasnoun{hull1999properties}. While the relaxation time can be
increased arbitrarily by setting the initial condition close to $T_u$,
we estimate the characteristic ``thermal memory'' timescale as the
maximum time it takes the middle ring to reach the stable steady
states when its starting temperature is taken to be that of either the
hot or cold resonators, which are 0.1s and 1s, respectively.

Compared to previous implementations based on phase-transition
materials,~\cite{dyakov2015near,kubytskyi2014radiative} the transition
times achieved here are of the same order of magnitude while the range
of operating temperatures is wider by an order of magnitude. While the
relaxation process can in principle be hastened by exploiting large
thermo-optic coefficients and/or larger $Q$, thus decreasing the
operating temperature range, the former are constrained by material
choices while the latter lead to decreased flux rates. Aside from
careful engineering of the coupling rates and resonator frequencies
needed to achieve bistability, a thermal memory based on this setup
requires good thermal insulation and suitable choice of materials
exhibiting large thermo-optic coefficients for speed and improved
stability (reliability).


\begin{figure}[t!]
\centering
\includegraphics[width=0.9\linewidth]{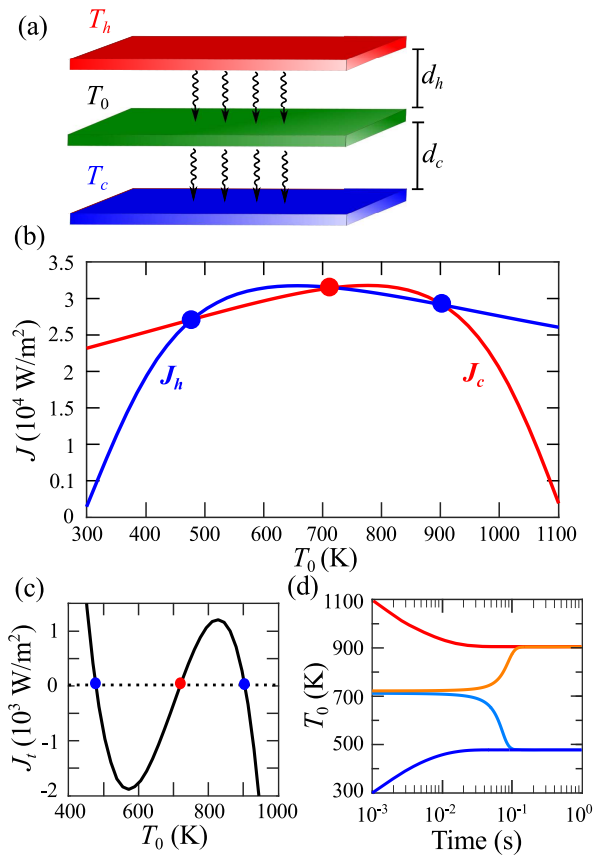} 
\caption{(a) Schematic showing system of three planar GaAs thin films
  of equal thickness $200$nm and separated by distances $d_h=48$nm and
  $d_c=45$nm, where the hot (red) and cold (blue) films are maintained
  at fixed temperatures, $T_h=1100$K and $T_c=300$K, while the middle
  film (green) is thermally insulated from the surroundings and has a
  variable $T_0$. (b) Heat flux rates per unit area, $J_h$ and $J_c$,
  and (c) net flux rate $J_t=J_h-J_c$, as a function of $T_0$. The
  nonmonotonicity in $J_t(T_0)$ results in three steady states, two of
  which are stable (blue dots) while the remaining one is unstable
  (red dot). (d) Temporal relaxation of the intermediate film starting
  from either $T_h$ (red) or $T_c$ (blue), along with the value of the
  unstable temperature point $T_u$, as it reaches the steady-state
  temperatures, $T_{s1}$ and $T_{s2}$.}
\label{fig:figGaAs}
\end{figure}

\emph{Thin films:} One possible way to increase the speed of such a
thermal memory device is to exploit planar polaritonic materials,
which offer enhanced heat flux rates owing to the large number of
surface localized resonances they can support. In what follows, we
consider one such example, shown schematically in \figref{figGaAs}(a),
consisting of three GaAs thin films exchanging heat radiatively in the
near field, where the hot and cold films are again held at fixed
temperatures $T_h$ and $T_c$, while the intermediate film is thermally
insulated from its surroudings and hence described by a variable
temperature $T_0$. Such a three-body planar configuration has been
studied previously using scattering formulations, with the various
flux rates obtained through straightforward calculation of the
reflection/transmission matrices in this
geometry,~\cite{messina2014three}, as described in detail
in~\citeasnoun{francoeur2009solution}. Here, we exploit this approach
to consider a full calculation of the flux rates that includes
thermo-optic effects in GaAs, obtained
from~\citeasnoun{blakemore1982semiconducting}, assuming operating
temperatures $T_h=1100$K and $T_c=300$K, 200nm films, and vacuum gaps
of $d_h=48$nm and $d_c=45$nm.


\Figref{figGaAs}(b) shows the computed flux rates per unit area, $J_h$
and $J_c$, as a function of $T_0$, while the net flux entering/leaving
the middle film $J_t$ is shown in \figref{figGaAs}(c). As before, the
thermo-optic induced NDTR results in two stable steady states at
$T_{s1}=430$K and $T_{s2}=900K$ (blue dots) along with an unstable
steady state at $T_u=710K$ (red dot). \Figref{figGaAs}(d) illustrates
the relaxation of $T_0$ from $T_h,T_c,T_{u}^+,T_u^-$ to the nearest
stable steady states, which is substantially decreased compared to
ring resonators due to the significantly larger flux rates $\gtrsim
10^4$W/m$^2$ attained in this setup. Moreover, the characteristic
timescale associated with such a relaxation, i.e. the maximum time it
takes the middle film to reach a steady state when starting from $T_h$
or $T_c$, is $\approx 0.1$s and can be further decreased by going to
smaller separations or smaller film thicknesses.

Note that a related NDTR-based mechanism was suggested in
\citeasnoun{zhu2012negative} in a system of two planar SiC plates. In
that work, the heat flux between the two plates was shown to vary
nonmonotonically at very high temperatures $T\sim 1500$K and small
separations $d \sim 15$nm, in which case application of a constant
(temperature independent) external flux leads to thermal
bistability. Another recent work proposed a nanothermomechanical
memory where NDTR is achieved at very high temperatures $T\sim 1100K$
by exploiting the nonmonotonic dependence of near field heat transfer
on the separation between two planar slabs, as actuated by the thermal
expansion of a mechanical support. The three-body system explored in
this work relaxes some of these stringent operating conditions,
allowing for a wide range of operating temperatures (steady state
temperatures $\lesssim 1000$K) and flux rates. Furthermore, our
proposed scheme also offers flexibility with respect to material
choices in that it does not rely on phase-change
materials~\cite{kubytskyi2014radiative} and could be realized with a
wide range of materials exhibiting strong thermo-optic effects, such
as chalcogenide glasses~\cite{seddon1995chalcogenide},
silica~\cite{palik1998handbook}, and silicon
carbide~\cite{palik1998handbook}, among many others. While thermal
memory devices based on phase-transition materials offer a smaller
operating temperature range (close to room temperature in case of
vanadium dioxide~\cite{dyakov2015near}) and have also been shown to
lead to multistability~\cite{ito2016multilevel}, our three-body scheme
leads to wider temperature differences between the steady states,
thereby guaranteeing stability against temperature and flux
perturbations.


\emph{Concluding remarks:} We demonstrated a simple scheme to realize
temperature bistability in all-passive systems comprising multiple
coupled resonant modes. We provided concrete predictions of expected
operating conditions (including transition times of several hundred
milliseconds) in realistic designs involving either suspended Si ring
resonators or GaAs thin films. Since the underlying mechanism is very
general and not restricted to the proposed implementations, one
possible direction forward could be to explore other geometries such
as nanobeam resonators~\cite{quan2011deterministic}, multilayered thin
films~\cite{boriskina2015enhancement}, nanostructured
materials~\cite{luo2013nanoscale,hafeli2011temperature} and different
choices of materials~\cite{seddon1995chalcogenide,palik1998handbook},
where one could potentially observe larger heat exchange. With rapidly
advancing nanotechnology, the understanding of this and related
thermal phenomena could be important for nanoscale heat management.

\emph{Acknowledgements:} We would like to thank Riccardo Messina and
Weiliang Jin for useful comments. This work was partially supported by
the National Science Foundation under Grant no. DMR-1454836 and by the
Princeton Center for Complex Materials, a MRSEC supported by NSF Grant
DMR 1420541.

\bibliographystyle{unsrt} \bibliography{photon}

\begin{thebibliography}{10}

\bibitem{cahill2003nanoscale}
David~G Cahill, Wayne~K Ford, Kenneth~E Goodson, Gerald~D Mahan, Arun Majumdar,
  Humphrey~J Maris, Roberto Merlin, and Simon~R Phillpot.
\newblock Nanoscale thermal transport.
\newblock {\em Journal of Applied Physics}, 93(2):793--818, 2003.

\bibitem{li2012colloquium}
Nianbei Li, Jie Ren, Lei Wang, Gang Zhang, Peter H{\"a}nggi, and Baowen Li.
\newblock Colloquium: Phononics: Manipulating heat flow with electronic analogs
  and beyond.
\newblock {\em Reviews of Modern Physics}, 84(3):1045, 2012.

\bibitem{song2015near}
Bai Song, Anthony Fiorino, Edgar Meyhofer, and Pramod Reddy.
\newblock Near-field radiative thermal transport: From theory to experiment.
\newblock {\em AIP Advances}, 5(5):053503, 2015.

\bibitem{otey2010thermal}
Clayton~R Otey, Wah~Tung Lau, Shanhui Fan, et~al.
\newblock Thermal rectification through vacuum.
\newblock {\em Physical Review Letters}, 104(15):154301, 2010.

\bibitem{wang2008thermal}
Lei Wang and Baowen Li.
\newblock Thermal memory: a storage of phononic information.
\newblock {\em Physical review letters}, 101(26):267203, 2008.

\bibitem{ben2014near}
Philippe Ben-Abdallah and Svend-Age Biehs.
\newblock Near-field thermal transistor.
\newblock {\em Physical review letters}, 112(4):044301, 2014.

\bibitem{sklan2015splash}
Sophia~R Sklan.
\newblock Splash, pop, sizzle: Information processing with phononic computing.
\newblock {\em AIP Advances}, 5(5):053302, 2015.

\bibitem{chang2007nanotube}
Chih-Wei Chang, D~Okawa, H~Garcia, A~Majumdar, and A~Zettl.
\newblock Nanotube phonon waveguide.
\newblock {\em Physical review letters}, 99(4):045901, 2007.

\bibitem{kubytskyi2014radiative}
Viacheslav Kubytskyi, Svend-Age Biehs, and Philippe Ben-Abdallah.
\newblock Radiative bistability and thermal memory.
\newblock {\em Physical review letters}, 113(7):074301, 2014.

\bibitem{dyakov2015near}
Sergey~A Dyakov, Jin Dai, Min Yan, and Min Qiu.
\newblock Near field thermal memory based on radiative phase bistability of
  vo2.
\newblock {\em Journal of Physics D: Applied Physics}, 48(30):305104, 2015.

\bibitem{zhu2012negative}
Linxiao Zhu, Clayton~R Otey, and Shanhui Fan.
\newblock Negative differential thermal conductance through vacuum.
\newblock {\em Applied Physics Letters}, 100(4):044104, 2012.

\bibitem{elzouka2014near}
Mahmoud Elzouka and Sidy Ndao.
\newblock Near-field nanothermomechanical memory.
\newblock {\em Applied Physics Letters}, 105(24):243510, 2014.

\bibitem{chen2014photon}
Zhen Chen, Carlaton Wong, Sean Lubner, Shannon Yee, John Miller, Wanyoung Jang,
  Corey Hardin, Anthony Fong, Javier~E Garay, and Chris Dames.
\newblock A photon thermal diode.
\newblock {\em Nature communications}, 5, 2014.

\bibitem{maghsoudi2016thermally}
Elham Maghsoudi and Michael~James Martin.
\newblock Thermally actuated buckling beam memory: a non-volatile memory
  configuration for extreme space exploration environments.
\newblock {\em Microsystem Technologies}, 22(5):1043--1053, 2016.

\bibitem{rueckes2000carbon}
Thomas Rueckes, Kyoungha Kim, Ernesto Joselevich, Greg~Y Tseng, Chin-Li Cheung,
  and Charles~M Lieber.
\newblock Carbon nanotube-based nonvolatile random access memory for molecular
  computing.
\newblock {\em science}, 289(5476):94--97, 2000.

\bibitem{nguyen1999radiation}
DN~Nguyen, SM~Guertin, GM~Swift, and AH~Johnston.
\newblock Radiation effects on advanced flash memories.
\newblock {\em IEEE Transactions on Nuclear Science}, 46(6):1744--1750, 1999.

\bibitem{bagatin2017space}
Marta Bagatin, Simone Gerardin, and Alessandro Paccagnella.
\newblock Space and terrestrial radiation effects in flash memories.
\newblock {\em Semiconductor Science and Technology}, 32(3):033003, 2017.

\bibitem{holman2001heat}
JP~Holman.
\newblock Heat transfer, eighth si metric edition, 2001.

\bibitem{rytov1988principles}
Sergei~M Rytov, Yurii~A Kravtsov, and Valeryan~I Tatarskii.
\newblock Principles of statistical radiophysics 2.
\newblock 1988.

\bibitem{zhang2013multibistability}
Libin Zhang, Yonghao Fei, Tongtong Cao, Yanmei Cao, Qingyang Xu, and Shaowu
  Chen.
\newblock Multibistability and self-pulsation in nonlinear high-q silicon
  microring resonators considering thermo-optical effect.
\newblock {\em Physical Review A}, 87(5):053805, 2013.

\bibitem{bogaerts2012silicon}
Wim Bogaerts, Peter De~Heyn, Thomas Van~Vaerenbergh, Katrien De~Vos, Shankar
  Kumar~Selvaraja, Tom Claes, Pieter Dumon, Peter Bienstman, Dries
  Van~Thourhout, and Roel Baets.
\newblock Silicon microring resonators.
\newblock {\em Laser \& Photonics Reviews}, 6(1):47--73, 2012.

\bibitem{xia2013suspended}
Yang Xia, Ciyuan Qiu, Xuezhi Zhang, Weilu Gao, Jie Shu, and Qianfan Xu.
\newblock Suspended si ring resonator for mid-ir application.
\newblock {\em Optics letters}, 38(7):1122--1124, 2013.

\bibitem{suh2004temporal}
Wonjoo Suh, Zheng Wang, and Shanhui Fan.
\newblock Temporal coupled-mode theory and the presence of non-orthogonal modes
  in lossless multimode cavities.
\newblock {\em IEEE Journal of Quantum Electronics}, 40(10):1511--1518, 2004.

\bibitem{khandekar2015thermal}
Chinmay Khandekar, Zin Lin, and Alejandro~W Rodriguez.
\newblock Thermal radiation from optically driven kerr ($\chi$ (3)) photonic
  cavities.
\newblock {\em Applied Physics Letters}, 106(15):151109, 2015.

\bibitem{zhu2013temporal}
Linxiao Zhu, Sunil Sandhu, Clayton Otey, Shanhui Fan, Michael~B Sinclair, and
  Ting~Shan Luk.
\newblock Temporal coupled mode theory for thermal emission from a single
  thermal emitter supporting either a single mode or an orthogonal set of
  modes.
\newblock {\em Applied Physics Letters}, 102(10):103104, 2013.

\bibitem{Hausbook}
Hermann~A. Haus.
\newblock {\em Waves and field in optoelectronics}.
\newblock Prentice Hall, 1984.

\bibitem{shaw2008fabrication}
Michael~J Shaw, Michael~R Watts, and Gregory~N Nielson.
\newblock Fabrication techniques for creating a thermally isolated tm-fpa
  (thermal microphotonic focal plane array).
\newblock In {\em MOEMS-MEMS 2008 Micro and Nanofabrication}, pages
  688308--688308. International Society for Optics and Photonics, 2008.

\bibitem{hull1999properties}
Robert Hull.
\newblock {\em Properties of crystalline silicon}.
\newblock Number~20. IET, 1999.

\bibitem{messina2014three}
Riccardo Messina and Mauro Antezza.
\newblock Three-body radiative heat transfer and casimir-lifshitz force out of
  thermal equilibrium for arbitrary bodies.
\newblock {\em Physical Review A}, 89(5):052104, 2014.

\bibitem{francoeur2009solution}
Mathieu Francoeur, M~Pinar Meng{\"u}{\c{c}}, and Rodolphe Vaillon.
\newblock Solution of near-field thermal radiation in one-dimensional layered
  media using dyadic green's functions and the scattering matrix method.
\newblock {\em Journal of Quantitative Spectroscopy and Radiative Transfer},
  110(18):2002--2018, 2009.

\bibitem{blakemore1982semiconducting}
JS~Blakemore.
\newblock Semiconducting and other major properties of gallium arsenide.
\newblock {\em Journal of Applied Physics}, 53(10):R123--R181, 1982.

\bibitem{seddon1995chalcogenide}
AB~Seddon.
\newblock Chalcogenide glasses: a review of their preparation, properties and
  applications.
\newblock {\em Journal of Non-Crystalline Solids}, 184:44--50, 1995.

\bibitem{palik1998handbook}
Edward~D Palik.
\newblock {\em Handbook of optical constants of solids}, volume~3.
\newblock Academic press, 1998.

\bibitem{ito2016multilevel}
Kota Ito, Kazutaka Nishikawa, and Hideo Iizuka.
\newblock Multilevel radiative thermal memory realized by the hysteretic
  metal-insulator transition of vanadium dioxide.
\newblock {\em Applied Physics Letters}, 108(5):053507, 2016.

\bibitem{quan2011deterministic}
Qimin Quan and Marko Loncar.
\newblock Deterministic design of wavelength scale, ultra-high q photonic
  crystal nanobeam cavities.
\newblock {\em Optics express}, 19(19):18529--18542, 2011.

\bibitem{boriskina2015enhancement}
Svetlana~V Boriskina, Jonathan~K Tong, Yi~Huang, Jiawei Zhou, Vazrik Chiloyan,
  and Gang Chen.
\newblock Enhancement and tunability of near-field radiative heat transfer
  mediated by surface plasmon polaritons in thin plasmonic films.
\newblock In {\em Photonics}, volume~2, pages 659--683. Multidisciplinary
  Digital Publishing Institute, 2015.

\bibitem{luo2013nanoscale}
Tengfei Luo and Gang Chen.
\newblock Nanoscale heat transfer--from computation to experiment.
\newblock {\em Physical Chemistry Chemical Physics}, 15(10):3389--3412, 2013.

\bibitem{hafeli2011temperature}
Andrew~K Hafeli, Eden Rephaeli, Shanhui Fan, David~G Cahill, and Thomas~E
  Tiwald.
\newblock Temperature dependence of surface phonon polaritons from a quartz
  grating.
\newblock {\em Journal of Applied Physics}, 110(4):043517, 2011.

\end{thebibliography}

\end{document}